# Fuzzy band structure of quantum dots by Bloch Orbital Expansion, unconventional insights into geometry-electronic structure relations

*Zeger Hens[1,2], Jordi Llusar[3], Ivan Infante[3,4]*

[1]Physics and Chemistry of Nanostructures, Ghent University, Gent, Belgium

[2]Center for Nano- and Biophotonics, Ghent University, Gent, Belgium

[3]BCMaterials, Basque Center for Materials, Applications, and Nanostructures, UPV/EHU Science Park, Leioa 48940, Spain

[4]Ikerbasque Basque Foundation for Science, Bilbao 48009, Spain

**Abstract.** The extension of ab-initio methods like density functional theory (DFT) to quantum dot (QD) geometries has enabled researchers to explore relationships between QD surface termination and electronic structure. However, fully utilizing the data from DFT requires novel classification methods for QD orbitals. Here, we identify relationships between QD geometry and electronic structure by transforming real-space QD orbitals into momentum-space using Bloch orbital expansion (BOE), yielding a fuzzy QD band structure. Comparing with bulk band structures, we show that truncated, unpassivated facets in III-V and II-VI QDs produce mid-gap orbitals derived from bulk surface orbitals; an identification challenging in real space. QDs with reconstructed facets, however, feature delocalized orbitals formed by superposition of bulk Bloch orbitals. Moreover, we prepare for the first time atomistic core/shell QD models and by the BOE expansion we can clearly identify the core/shell band alignment not possible in real space. These findings emphasize BOE as a vital tool for connecting computational and experimental insights in nanocrystal research.

Quantum dots (QDs) are a state of matter that exists between bulk semiconductors and single molecules.[1] In QDs, atoms form a regular lattice similar to bulk crystals, yet the QD electronic structure features discrete, delocalized states that resemble molecular orbitals (MOs). Especially through colloidal synthesis, a wide variety of semiconductors can be synthesized as QDs with varying size and shape or involving intricate heterostructures.[2] Research on colloidal QDs has greatly benefited from computational and theoretical analysis. Typical solid-state physics methods, such as effective mass or $k \cdot p$ theory,[3] atomistic tight binding,[4] or pseudopotential methods,[5] yield single-electron QD orbitals as linear combinations of bulk Bloch orbitals that exhibit envelope symmetries similar to atomic hydrogen. Moreover, by promoting electrons to unoccupied orbitals, excited states can be constructed as interacting electron-hole pairs, i.e., (multi)excitons, in good agreement with experimental findings.[6, 7] However, these approaches are less suited to analyze surface-related properties, such as the reconstruction or ligand-passivation of QD facets or predict band offsets within a heterostructure.

Opposite from solid-state physics methods, density-functional theory (DFT) can provide MOs as a linear combination of atomic orbitals, calculated directly from atomistic QD models that include a realistic description of the QD surface.[8-12] Using DFT, QD structures are obtained through energy and structural minimization, and different surface terminations can be explored with MOs that are not predesigned as consisting of bulk Bloch orbitals. Despite limitations in accurately describing bandgaps and excited-state geometries of QDs,[13] DFT has been proven valuable to identify trap states in QDs,[8, 9, 14-17] describe excited states carrier dynamics,[18, 19] determine ligand-QD binding energies,[20, 21] and calibrate force fields for molecular dynamics simulations.[22-28] However, QD models with realistic surface termination often feature frontier orbitals dominated by surface contributions.[29] This predominance complicates the identification of delocalized orbitals that resemble the bulk Bloch orbitals observed in solid-state calculations of the same materials. Bridging this gap between DFT and solid-state physics methods would create an entirely new perspective for the computational study of semiconductor QDs, especially for the traditional II-VI, IV-VI, III-V and metal halide perovskite semiconductor families, with ramifications well beyond the case of semiconductor nanocrystals.

Here, we show that projection on Bloch orbitals provides a unique tool to classify QD orbitals obtained through DFT as derived from bulk Bloch or surface orbitals. Building on the concept of a fuzzy QD band structure for silicon-based nanocrystals,[30, 31] we confirm that this so-called Bloch Orbital Expansion (BOE) transforms the QD MOs into a fuzzy band structure that mimics the semiconductor bulk bands. For a wide array of QDs, we use BOE to distinguish surface-localized and delocalized frontier orbitals and identify the dominant symmetry of delocalized orbitals. Next, we implement BOE to highlight that surface reconstructions are key to suppress the admixing of bulk surface orbitals in QD MOs and obtain delocalized frontier orbitals. We then extend BOE to core/shell QDs, which results in a first method to predict band offsets; a heterostructure characteristic that other computational methods take as an adjustable parameter. We conclude that BOE provides a broad and exceptionally insightful perspective on the QD electronic structure, in particular for relating size quantization to the atomic structure of QDs, including realistic surface modifications.

**Results.**

### Bloch Orbital Expansion of Single-Electron States

According to the Bloch theorem, an orbital $\psi_k(x)$ on an infinite periodic lattice can be expressed as the product of a plane wave – characterized in 1 dimension by the wavenumber $k$ – and a part $u_k(x)$ with the periodicity of the lattice:

$$\psi_k(x) = e^{ikx} u_k(x) \quad (1)$$

Within a QD, orbitals $\varphi_n(x)$ can be written as a linear combination of such Bloch orbitals at constant energy:

$$\varphi_n(x) = \sum_k c_{k,n} u_k(x) e^{ik\cdot x} = \sum_{k,m} c_{k,n} u_{k,m} e^{i\left(\frac{2\pi}{a}m+k\right)\cdot x} \quad (2)$$

In the last step, $u_k(x)$ was expanded as a Fourier series, which assigns the Fourier coefficients $u_{k,m}$ to each Brillouin zone (BZ) $m$ of the reciprocal lattice.

By Bloch-orbital expansion (BOE), a single electron state is represented through the expansion coefficients $c_{k,n}$ rather than the orbital $\varphi_n(x)$. As outlined by Hapala et al,[30] BOE can be achieved by considering the Fourier transform $\Phi_n(k)$ of $\varphi_n(x)$. Fourier transformation turns a Bloch orbital $\psi_k(x)$ into a comb of delta functions, one in each reciprocal lattice cell. Assuming $u_k(x)$ is normalized, the sum $S_k$ of the amplitude squared of the weight of the delta functions of a single comb reads:

$$S_{k,n} = \sum_m c_{k,n}^* c_{k,n} u_{k,m}^* u_{k,m} = c_{k,n}^* c_{k,n} \sum_m u_{k,m}^* u_{k,m} = c_{k,n}^* c_{k,n} \quad (3)$$

Since $c_{k,n}^* c_{k,n}$ provides the weight of a given Bloch orbital in the expansion of $\varphi_n(x)$, BOE can be accomplished by folding back $\Phi_n(k)^*\Phi_n(k)$ to the first BZ (BZ1). Note that by the finite size of a QD, $\Phi_n(k)$ will be smeared out in reciprocal space rather than being a set of delta functions, hence the notion of a fuzzy band structure.

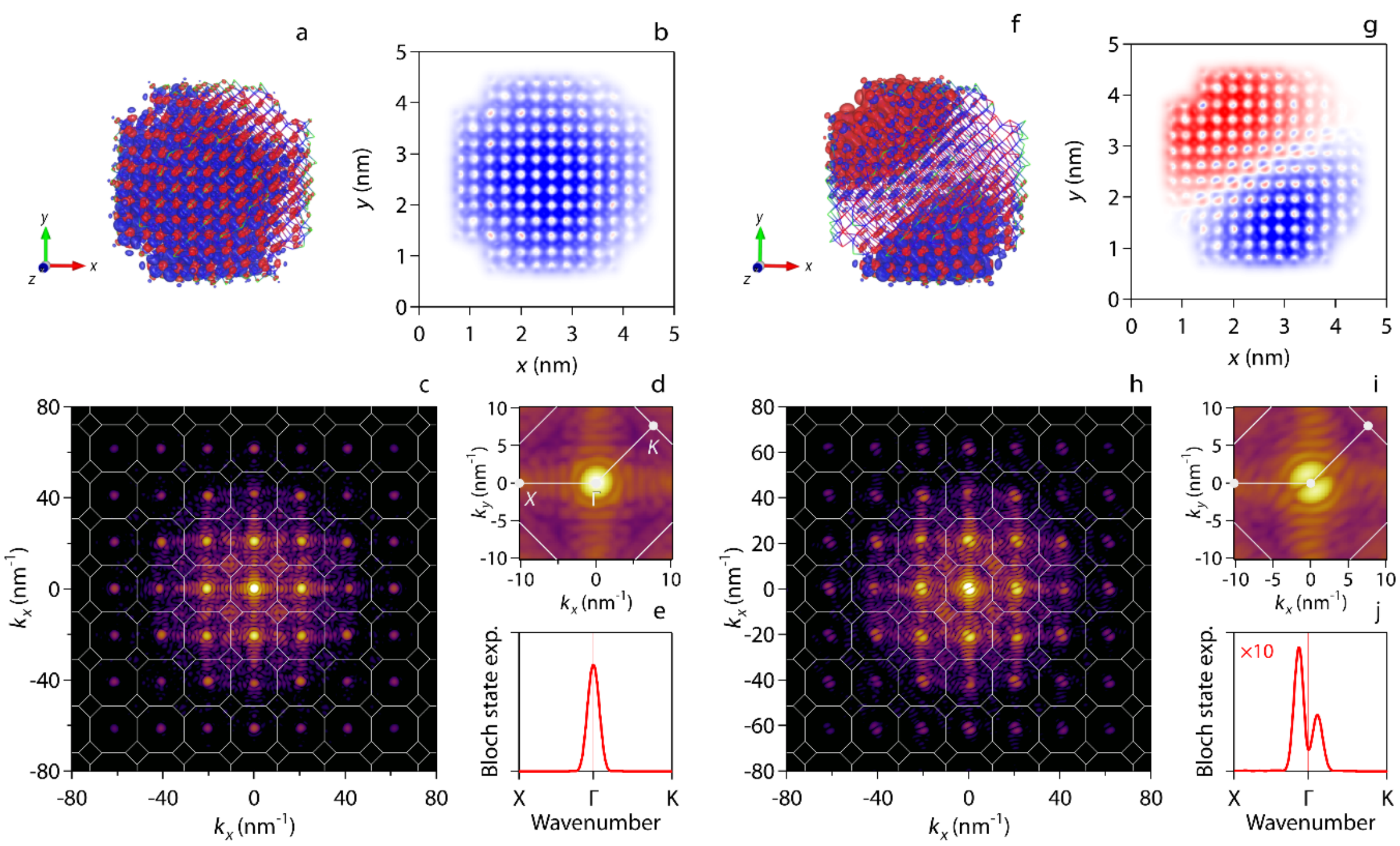

**Figure 1.** Bloch orbital expansion of QD orbitals. (a) Isosurface of the LUMO of a 3.5 nm Cl-terminated CdSe QD, highlighting positive and negative parts in red and blue, respectively. (b) Projection of the LUMO on the $xy$ plane, using (positive) red and (negative) blue shading to represent the magnitude of the wavefunction. (c) Fourier transform of the wavefunction in the $k_x k_y$ plane, as calculated from the projected orbital. The white line pattern represents the intersection of the $k_x k_y$ plane with the respective BZs. (d) Fourier transform folded back on BZ1, with an indication of a path through BZ1 along which the BOE can be calculated more efficiently. (e) Representation of the BOE along the identified path in BZ1. (f-j) The same, for LUMO+1. Note how the nodal plane of the wavefunction envelope in real space results in a splitting of the BOE relative to $\Gamma = 0$. Note that the vertical axis in Figure 1j is expanded 10-fold relative to Figure 1e.

**Figure 1** illustrates the approach by means of the LUMO and LUMO+1 of a chloride-terminated 3.5 nm CdSe QD, which have envelopes with S and P symmetry, respectively. Details of this 1259-atom QD model will be discussed later. By projecting the wavefunction on the $xy$ plane, one obtains $\Phi_n(k)$ in the $k_x k_y$ plane shown in **Figure 1c** and **1h**. As expected, $\Phi_n(k)$ exhibits a patterning that tracks the BZs of the reciprocal lattice, and that can be folded back to BZ1 as shown in **Figure 1d** and **1i**. In practice, we calculated this folded Fourier transform along a 1D path through BZ1, see **Supplemental Information S1**. This approach massively reduces the computational cost of the Fourier transform and enables us to move with high resolution along $k$ points of interest, in comparison with bulk band structures. As shown in **Figure 1e** and **1j**, the result is a 1D representation of the BOE along the path through BZ1. Note that, even if the path is not fully aligned with the shape of $\Phi_n(k)$, BOE gives direct access to the envelope symmetry, with the S envelope yielding a BOE centered at $\Gamma = 0$, and the P envelope split in two parts in agreement with the real space wavefunction having a nodal plane.

### From Bloch Orbital Expansion to Fuzzy Band Structure

**Figure 2** displays the outcome of a BOE, applied to a 4 nm Pb706S586Cl240 QD (PbS-1532), the relaxed structure of which is shown in **Figure 2a**. This QD is cation rich, in agreement with experiments,[32] and exposes (100) and (111) facets. Emulating X-type ligands, Cl is bound to the (111) facets. **Figure 2c** represents the BOE calculated along the path indicated in **Figure 2b** as an energy vs. wavenumber image plot, see **Supplemental Information S2** for computational details. What emerges through this representation is a fuzzy band structure of PbS-1532. The direct comparison of this pattern with the bulk PbS energy bands calculated at the same level of theory, see **Supplemental Information S3**, leads to multiple observations:

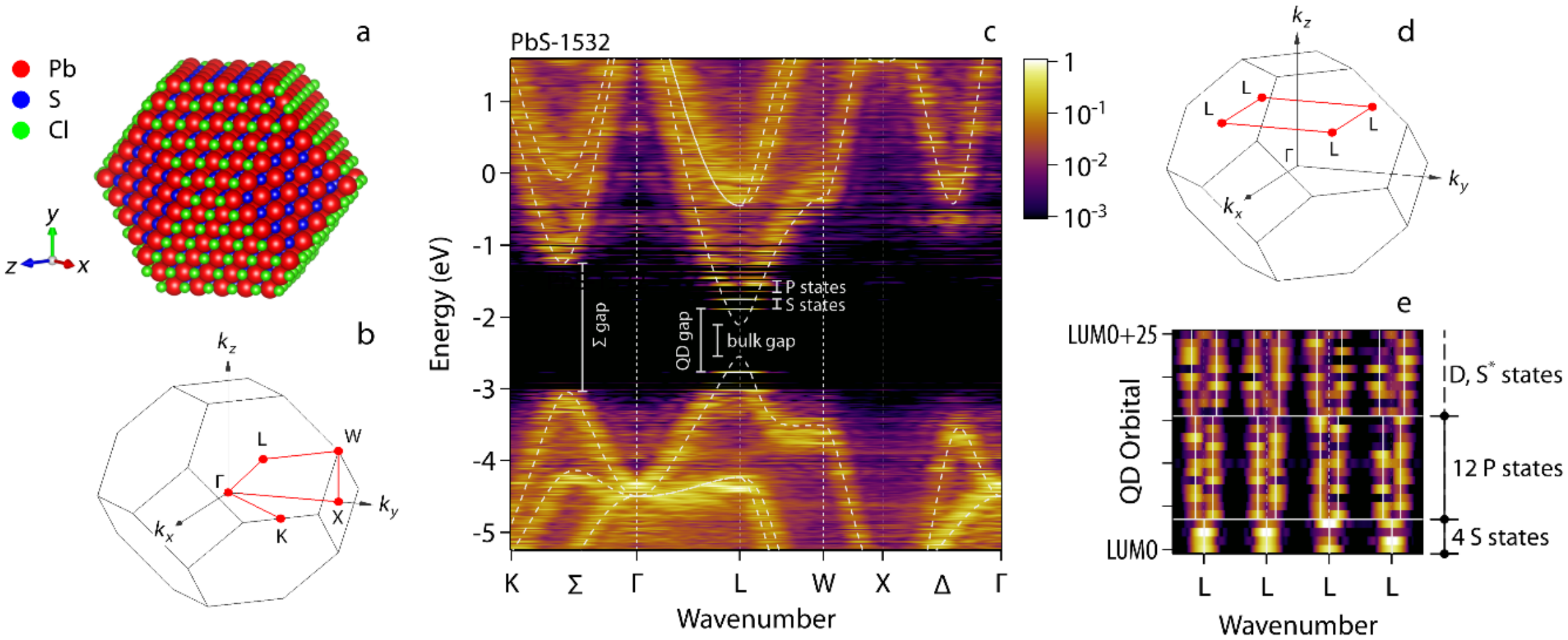

**Figure 2.** The fuzzy band structure of a 4 nm $Pb_{706}S_{586}Cl_{240}$ (PbS-1532) QD. (a) Representation of the atomic structure of PbS-1532. (b) fcc Brillouin zone, with the path along which the BOE is calculated indicated in red. (c) Color-coded image representation of the BOE as a function of the wavenumber along the path in BZ1 and the energy of the different orbitals. Color coding in logarithmic scale as indicated. For obtaining a consistent color coded fuzzy band structure, the BOE has been accumulated in energy bins 12.5 meV wide. (d) Indication of the path followed to probe orbitals having a BOE around one of the 4 equivalent L points. (e) Color coded image representation of the BOE of the LUMO to LUMO+25 as a function of the QD orbital number and the wavenumber. Horizontal lines separate sets of orbitals with similar envelope symmetry, vertical lines indicate the local maxima of the BOE of each set of orbitals.

- The absence of significant quantization around the van 't Hoff singularity along the Σ direction (Γ − K) enables the bulk and the PbS-1532 fuzzy bands to be aligned, which attests to the close agreement between both sets of bands.
- The frontier orbitals of PbS-1532 develop around the L point, similar to the bulk band gap. The energy difference of both the HOMO and the LUMO with the bulk band edges amounts to 0.22 eV. While these shifts underestimate the actual quantization energy in PbS QDs – a known limitation of DFT – their similarity is consistent with the comparable effective masses of electrons and holes.

- In both bands, a sequence of orbitals with S and P envelopes can be identified, either as having a maximal BOE at L, or a fixed splitting in two lobes around L.

Interestingly, PbS has a 4-fold degenerate band gap at 4 equivalent L points. In **Figure 2d-e**, we explore the impact of this property on the CB eigenstates using BOE along a path that addresses these different L points. As highlighted in **Figure 2e**, 4 states with an S-type and 12 states with a P-type envelope can be distinguished – in agreement with quantization in 4 band-edge minima.[33, 34] Moreover, these states feature contributions from multiple L points – reflecting intervalley coupling – with the most delocalized S-state corresponding to the LUMO. Higher lying states can be interpreted as having D- and S*-type envelopes, characterized either by two lobes with a wider splitting than P envelopes or an additional central lobe at $k = \mathrm{L}$. As outlined in **Supplementary Information S2**, a similar analysis of the VB eigenstates also yields 16 states with S- and P-symmetry, yet with more significant S-P coupling. Given this agreement between DFT orbitals and the understanding of QD orbitals according to solid-state physics methods,[35] we conclude that BOE of DFT-based QD orbitals creates a fuzzy QD band structure that provides a direct connection between these distinct computational approaches to QDs.

**Identification of surface-localized orbitals**

**Figure 3** displays results obtained on two InAs QDs, labeled InAs-1116 (3.0 nm) and InAs-2839 (4.2 nm), that pertain to a more extensive size series, see **Supplemental Information S4**. As indicated in **Figure 3a**, these QDs are In-rich, terminated by (100), In(111) and P(-111) facets and charge-compensated by Cl atoms bound to (100) and (111) facets. According to **Figure 3c-d**, BOE again produces a fuzzy band structure that aligns with the corresponding bulk bands. Both sets of bands were aligned through the high-energy $\Gamma_{15}$ state, see **Supplemental Information S4**. Comparing these QD fuzzy bands, the reduced broadening for InAs-2839 stands out, which is part of a systematic trend. As shown in **Figure 3e**, the LUMO of similar InAs QDs of different sizes has a BOE that peaks around $k = \Gamma$ – reflecting the S-type character of the LUMO – and narrows down with increasing size. Interestingly, for the different LUMOs, the width of the BOE compares well with the Fourier transform of a sphere having the QD size; a result underscoring the full wavefunction delocalization in these states.

Above the CB edge, the BOE features a set of discrete states, which have S (LUMO) or P-type (LUMO+1 – LUMO+3) character according to the shape of the BOE and the wavefunction, see **Figure 3h** and **3j**. Higher energy states, such as LUMO+4 are accordingly identified as having D-type symmetry, which confirms the interpretation of these unoccupied MOs as linear combinations of Bloch orbitals. The energy difference between the LUMO and the CB edge is the quantization energy, which reduces with QD size and exceeds the splitting between the LUMO and the P-state manifold by ~40%. Similar results have been obtained by tight-binding and pseudopotential calculations.[36, 37]

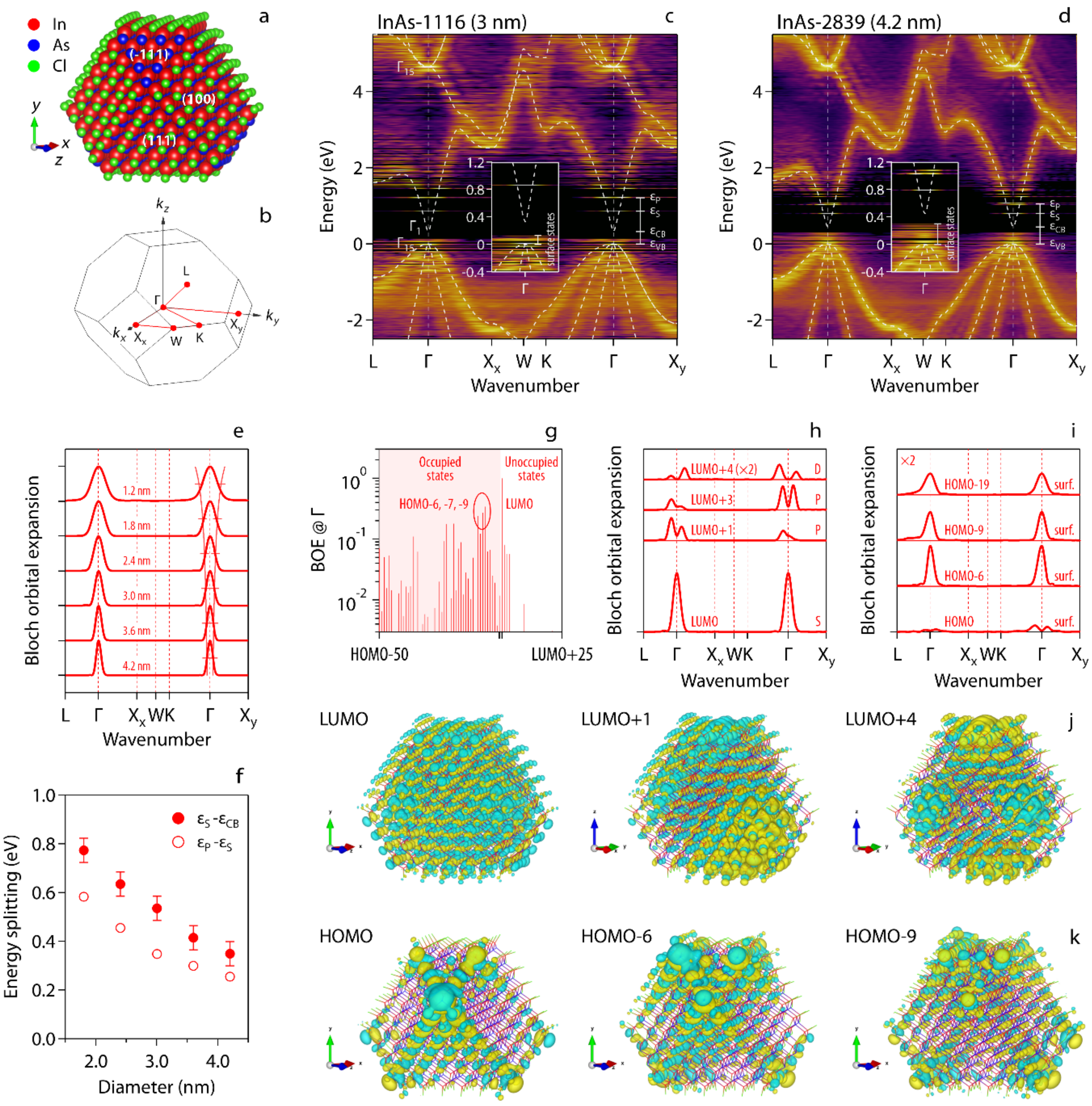

**Figure 3.** The fuzzy band structure of Cl-terminated InAs QDs. (a) Representation of the atomic structure of InAs-1116. (b) fcc Brillouin zone, with the path along which the BOE is calculated indicated in red. (c) Color coded image representation of the BOE of InAs-1116 (3.0 nm) as a function of the wavenumber along the path in BZ1 and the energy of the different states. Color coding similar to Figure 2c. For obtaining a consistent color coded fuzzy band structure, the BOE has been accumulated in energy bins 12.5 meV wide. (d) The same for InAs-2839 (4.2 nm). (e) BOE of the LUMO for the different InAs QDs analyzed, highlighting the S-type character of the orbital and the reduction of BOE broadening with increasing QD size. The light red line represents the full-width-

at-half-maximum for an envelope wavefunction that is constant across the QD volume. (f) Energy splitting between (filled markers) the S-state (LUMO) and the CB edge, and (open circles) the manifold of P-states and the S-state as a function of InAs QD size. (g) Overview of the BOE at the center of BZ1 (Γ point) for a selection of MOs near the VB- and CB edge. (h) BOE along the path in BZ1 for a selected set of unoccupied MOs. The envelope symmetry of the different orbitals is indicated. (i) BOE along the path in BZ1 for a selected set of occupied orbitals. (j) Isosurface of three unoccupied orbitals, featuring (LUMO) S, (LUMO+1) P and (LUMO+4) D-type envelope symmetry, respectively. (k) Isosurface of three occupied orbitals, showing pronounced facet and edge localization.

In stark contrast with the CB edge, several states of InAs-1116 and InAs-2839 appear above the bulk VB edge, see **Figure 3c-d**. Such in-gap states are incompatible with bulk Bloch orbitals, and must be related to local defects or consist of so-called bulk surface orbitals. These latter are exponentially decaying orbitals that solve the bulk Schrödinger equation but do not satisfy the infinite lattice boundary conditions.[38] Unlike PbS, tetrahedral semiconductor surfaces allow bulk surface orbitals to appear within the band gap.[39] As shown in **Figure 3g/3i**, the BOE of quite a few orbitals near the VB edge, such as HOMO-6, -7 and -9, peaks at $k = \Gamma$. While seemingly reflecting S-type symmetry, the isosurfaces of these MOs have dominant contributions from surface, edge and corner atoms that reminisce bulk surface orbitals, see **Figure 3j**. According to **Figure 3g**, other occupied MOs exhibit a smaller BOE at $k = \Gamma$, indicating the absence of S-type delocalized MOs around the VB edge. Instead, the noisier BOE and the selected isosurfaces suggest that occupied MOs in these QDs are best described as linear combinations of bulk Bloch and surface orbitals. Similar results were obtained on InP QDs, see **Supplementary Information S5**.

### The impact of surface reconstruction

**Figure 3k** makes clear that the in-gap orbitals of InAs-1116 mostly involve P-rich (-111) facets. In principle, similar orbitals can develop on In-rich (111) facets, but full Cl-passivation seemingly prevents such MOs from forming. This situation differs from CdSe QDs, where a smaller amount of Cl suffices to compensate for the Cd excess. As shown for a CdSe-771 model QD, similarly cut along (100), Cd(111) and Se(-111) facets and displayed in **Figure 4a**, the Cd(111) facet therefore features little bound Cl atoms. As outlined in **Figure 4b-g**, this limited passivation severely impacts on the electronic structure. Most strikingly, the BOE of the CdSe-771 states shows a multitude of in-gap orbitals, while MOs within the bulk bands only yield poorly distinguishable fuzzy bands. In addition, orbitals lack a well-defined envelope symmetry. This point is most clearly exemplified by those orbitals with the highest BOE at $k = \Gamma$, which can reflect an S-type envelope. However, **Figure 4d-g** demonstrates that even these orbitals have an ill-defined BOE and are mostly localized on (-111) and (111) facets in the respective case of occupied or unoccupied MOs. It thus appears that poorly passivated, bulk-like facets cause severe admixing of bulk surface orbitals in most – if not all – QD MOs and prevent the formation of delocalized orbitals as linear combinations of Bloch orbitals, a defining feature of strong quantization.

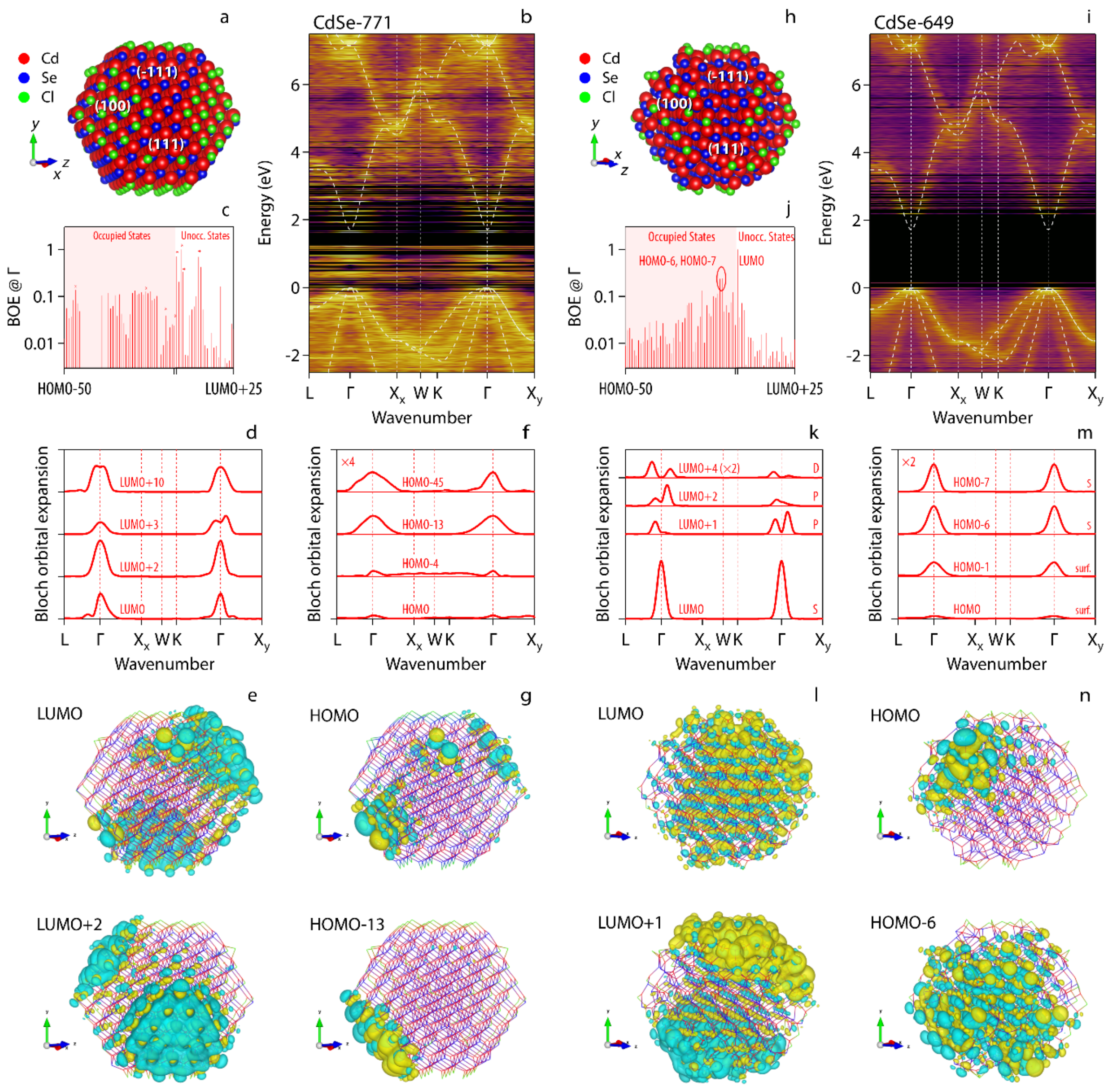

**Figure 4.** Impact on surface reconstruction on the QD eigenstates. (a) Representation of the atomic structure of CdSe-771, a QD model cut from bulk CdSe without surface reconstruction. (b) Color coded image representation of the BOE of CdSe-771 as a function of the wavenumber. The path in BZ1 and the color coding are similar to Figure 3c. For obtaining a consistent color coded fuzzy band structure, the BOE has been accumulated in energy bins 12.5 meV wide. (c) BOE at $k = \Gamma$ for occupied and unoccupied states near the band edges. States indicated with * are further analyzed in Figure 4d and 4f. (d) BOE along the path in BZ1 for a selected set of unoccupied levels. No envelope symmetry is assigned. (e) Isosurface representation of the wavefunction of LUMO and LUMO+2. (f) BOE along the path in BZ1 for a selected set of occupied levels. No envelope symmetry is assigned. (g) Isosurface representation of the wavefunction of HOMO and HOMO-13. (h-n) The same for CdSe-649, a QD model obtained from CdSe-771 through surface reconstructions involving atom removal. In Figure 4k and 4m, envelope symmetries have been indicated alongside the BOE of selected states.

Building on recent work that underscored the role of surface reconstructions in preventing bulk surface orbitals from developing in-gap QD MOs,[29] **Figure 4h-n** summarizes results obtained on a CdSe-649 model QD. Shown in **Figure 4h**, this structure was formed by removing surface atoms from CdSe-771, in line with known rules for reconstruction of polar facets of tetrahedral semiconductors,[40] see **Supplementary Information S6** for details. While the resulting surface reconstruction makes the layer of surface atom seem disordered, a BOE shows that admixing of bulk surface orbitals in the MOs of CdSe-649 is strongly suppressed, see **Figure 4i**. A well-defined fuzzy band structure emerges, no in-gap orbitals remain near the CB edge and only a few are present near the VB edge. Moreover, as shown in **Figure 4k-l**, the unoccupied orbitals exhibit again the expected sequence of states with S, P and D envelope symmetry, while HOMO-6 and HOMO-7 appear, according to the BOE profile and the wavefunction isosurface, as delocalized orbitals with an S-type envelope, see **Figure 4m-n**. We thus conclude that a reconstructed Cd(111) facet suppresses admixing of bulk surface orbitals in the QD unoccupied MOs, similar to full Cl passivation in the case of InAs and InP, while a reconstructed Se(-111) facet enables delocalized occupied orbitals with a well-defined envelope symmetry to be formed. Even so, the residual in-gap orbitals indicate that further reconstructions of the Se(-111) facet may be needed to entirely suppress admixing of bulk surface orbitals in the CdSe-649 MOs.

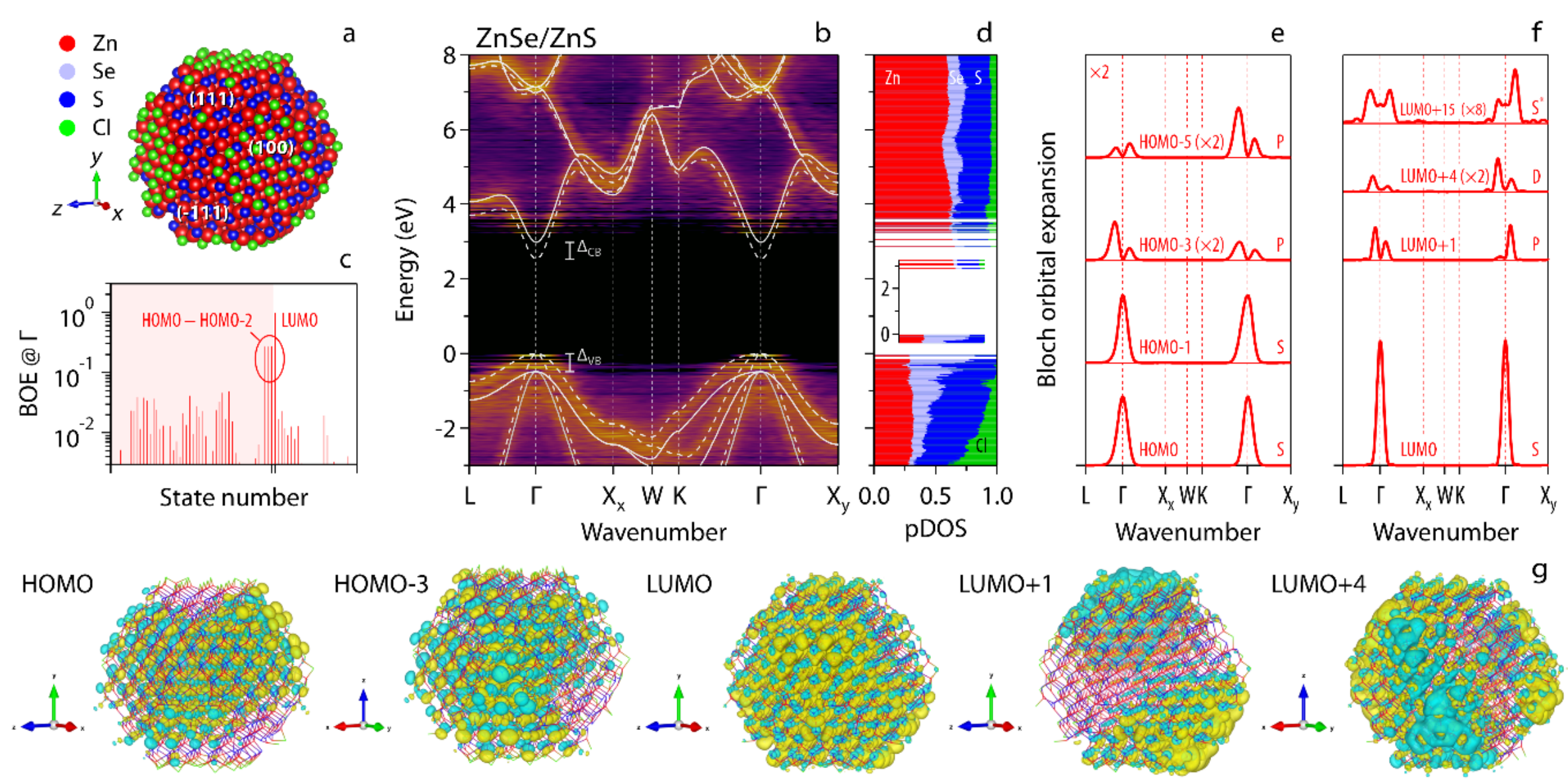

**Figure 5.** Fuzzy band structure of core/shell QDs. (a) Representation of the atomic structure of ZnSe/ZnS-1259. (b) Color coded image representation of the BOE of ZnSe/ZnS-1259 as a function of the wavenumber. The path in BZ1 and the color coding are similar to Figure 3c. For obtaining a consistent color coded fuzzy band structure, the BOE has been accumulated in energy bins 12.5 meV wide. The valence- and conduction-band offset has been indicated. (c) BOE at $k = \Gamma$ for occupied and unoccupied states near the band edges. (d) Partial density of states of the ZnSe/ZnS-1259 states, binned in similar 12.5 meV wide energy intervals. (e-f) BOE along the path in BZ1 for a selected set of occupied and unoccupied levels, with an indication of the corresponding envelope symmetry. (g) Isosurface representation of the wavefunction of selected occupied and unoccupied states of ZnSe/ZnS-1259.

## Band alignment in core/shell heterostructures

As a next step, we applied BOE to heteronanostructures. As outlined in **Supplemental Information S7** and **S8**, we used a generic, 1259 atom, II-VI core/shell model with reconstructed (100), (111) and (-111) facets. **Figure 5a** displays the case of ZnSe/ZnS-1259, for which **Figure 5b** represents a BOE energy/wavenumber image. Interestingly, a single fuzzy band structure is seemingly obtained, yet the combined ZnSe and ZnS bulk energy bands are required to connect the fuzzy bands of ZnSe/ZnS-1259 to bulk Bloch orbitals, see **Supplementary Information S8** for details. Such a combined description provides a direct estimate of the alignment between the core and shell bulk energy bands, which yields a straddling, so-called type I alignment for ZnSe/ZnS-1259, see **Figure 5b** and **Table 1**. Obtaining such figures for an actual QD structure rather than bulk lattices is a unique result that no other computations on QDs have provided.

Next to band alignment, BOE provides insight in the characteristics of the MOs of ZnSe/ZnS-1259. While plain ZnSe and ZnS-1259 QDs still show admixing of bulk surface orbitals in the occupied MOs, see **Supplementary Information S7**, the BOE of ZnSe/ZnS-1259 at $k = \Gamma$ points towards 3 highest occupied orbitals of S-type symmetry, the expected result given the 3-fold degeneracy of the DFT valence bands at $k = \Gamma$. Confirming this identification, the partial density of states shows a dominant contribution from the ZnSe core, not the ZnS surface, while the BOE and the wavefunction isosurface depict these orbitals as combining bonding atomic p orbitals with an S-type envelope, see **Figure 5d-f**. Furthermore, deeper occupied orbitals with P symmetry can be identified, see HOMO-3 in **Figure 5e-f**. Similarly, the unoccupied orbitals involve anti-bonding atomic s orbitals, with well-defined envelope symmetries extending to S*, see LUMO+15 in **Figure 5f-g** and **Supplementary Information S8**. Hence, BOE shows that shells with reconstructed facets strongly promote the formation of QD orbitals with well-defined envelope symmetry and multiplicity out of bulk Bloch orbitals.

**Table 1.** Overview of valence-band offset in eV as obtained from (bold) analysis of fuzzy bands obtained through Bloch-state expansion and (normal) literature-reported DFT calculations on bulk semiconductors.[41] The difference between both offsets is indicated in italics, highlighting significant (blue) positive and (red) negative deviations between QD and bulk DFT calculations.

| Core ↓ \| Shell → | CdS | ZnSe | ZnS |
|---|---|---|---|
| CdSe | **0.47**<br>0.42<br>*0.05* | **0.07**<br>0.07<br>*0.00* | **0.83**<br>0.6<br>***0.23*** |
| CdS | | **-0.60**<br>-0.35<br>***-0.25*** | **0.16**<br>0.18<br>*0.02* |
| ZnSe | **0.55**<br>0.35<br>***0.2*** | | **0.48**<br>0.53<br>*0.05* |

**Supplementary Information S8** provides an overview of a similar analysis on 6 more core/shell structures combining CdSe, CdS, ZnSe and ZnS. While all these systems feature orbitals with pronounced envelope symmetries, CdSe/CdS-1259 still shows a set of surface localized highest occupied orbitals that have a dominant contribution for atomic orbitals of sulfur. Hence, high band-gap shells do not necessarily avoid all admixing of bulk surface states. **Table 1** summarizes the resulting VB alignment of the core and shell bulk bands. These numbers confirm the common identification of these core/shell QDs as type I (CdSe/ZnS, ZnSe/ZnS), type II (CdS/ZnSe, ZnSe/CdS) or quasi-type II (CdSe/CdS, CdSe/ZnSe, CdS/ZnS). In addition, band alignments agree with published bulk calculations for common-anion and common-cation systems,[41] suggesting that band offsets are independent of the QD surface termination. However, these offsets do not add up to the offset in structures without common elements, such as CdSe/ZnS. Given the opposite deviation for CdS/ZnSe and ZnSe/CdS, this lack of transitivity could be related to the formation of an additional interfacial dipole in these structures.

## Discussion

By the specific treatment of the QD surface, solid-state physics methods are pre-designed for obtaining QD eigenstates as linear combinations of bulk Bloch orbitals. Using such delocalized orbitals, various opto-electronic properties, such as sizing curves,[32, 42] energy-level splitting,[43] symmetry-based selection rules and oscillator strengths of electronic transitions,[32, 44, 45] and exciton fine structure,[46-48] have been successfully described. DFT, on the other hand, enables for a more realistic implementation of surface termination. However, results often show a collapse of the QD band gap, and the appearance of surface-localized frontier orbitals. By transforming eigenstates into fuzzy QD energy bands that reproduce the bulk bands, BOE indicates that such results are not flaws of DFT, but rather result from the admixing of bulk surface orbitals into QD MOs. The relationship between surface reconstruction and electronic structure shows that this admixing strongly depends on the specific QD model used. With the fuzzy QD band structure obtained via BOE, meaningful QD models can be selected, QD states identified and key results – such as energy gaps and envelope symmetries – compared with predictions from solid-state physics methods. Moreover, also core/shell heterostructures exhibit fuzzy bands that can be described by the combined energy bands of two compounds, a finding that allows bulk semiconductor parameters, such as band offsets, to be derived from DFT calculations on realistic QD models.

While this work focused on II-VI, III-V and IV-VI semiconductors, BOE is not limited to these materials. Leveraging the ab initio nature of DFT, BOE can be applied to states of metal halide perovskites or chalcopyrite QDs, provided the correct folding schemes in reciprocal space are implemented. In addition, BOE enables QD models for more involved heterostructures, such as InAs/ZnSe, InP/ZnSe or CuInS2/ZnS to be assessed. Moreover, the close interplay observed between the QD surface and the characteristics of the QD MOS as obtained from DFT, is likely a general feature of nanocrystals. Given the increasing use of DFT as a computational approach to study nanocrystals, evaluating the characteristics of calculated orbitals – particularly the frontier orbitals – with respect to bulk energy bands appears an indispensable part of such studies.

## Methods

### Density functional theory calculations

Atomistic simulations at the Density Functional Theory (DFT) level have been carried out using the PBE (PbS),[49] or HLE17 (InAs, InP, CdSe, CdS, ZnSe, ZnS) exchange-correlation functional,[50] a double-ζ basis set (DZVP),[51] and the Gaussian and augmented plane waves method as implemented in the CP2k 2024.1 quantum chemical package.[52] The functionals were selected so as to obtain a reasonable estimate for the bulk band gap and the lattice parameter. Relativistic effects for all atoms are included using effective core-potentials. All structures have been optimized in the gas phase in cubic boxes that leave at least 6 Å of vacuum between the QD and the boundary of the box.

**Bloch orbital expansion**

Bloch orbital expansion is accomplished by folding back a 1D Fourier transform to a predefined path of connected line segments in the first Brillouin zone. To reduce the dimensionality from 3D to 1D, a 3D orbital is projected on a bundle of lines, where each line coincides with a given line segment in one of the Brillouin zones, and the Fourier transform of this 1D projected orbital is calculated by numerical integration. For Bloch orbital expansion, orbitals were printed as cube files using a grid spacing of $0.3a_0$ (Stride 2 2 2 in CP2K), with $a_0$ the Bohr radius. More details on the implemented procedure can be found in **Supplemental Information S1**. A free of charge and open-source code to compute fuzzy band structures from DFT calculations can be found on the GitHub repository: https://github.com/nlesc-nano/FuzzyQD

## ASSOCIATED CONTENT

**Supporting Information.**

S1. Theory, Implementation and Examples of Bloch Orbital Expansion

S2. Bloch Orbital Expansio of PbS Quantum Dot Models

S3. Calculation of Bulk energy bands

S4. Bloch Orbital Expansion of InAs Quantum Dot Models

S5. Bloch Orbital Expansion of InP Quantum Dot Models

S6. Bloch Orbital Expansion of Surface Reconstructed and Unreconstructed CdSe QD Models

S7. Bloch Orbital Expansion on Generic II-VI QD Models

S8. Bloch Orbital Expansion on core-shell II-VI@II-VI QD Models

AUTHOR INFORMATION

**Corresponding Authors**

*zeger.hens@ugent.be

*ivan.infante@bcmaterials.net ;

**Author Contributions**

The manuscript was written through contributions of all authors. All authors have given approval to the final version of the manuscript.

**Funding Sources**

FWO-Vlaanderen FWO-Vlaanderen (research project G0B2921N, Sabbatical bench fee K800324N), Ghent University (BOF-GOA 01G02124), Horizon Europe EIC Pathfinder program through project 101098649–UNICORN and IKUR strategy on behalf of the Department of Education of the Basque Government.

**ACKNOWLEDGMENTS**

ZH acknowledges the FWO-Vlaanderen (research project G0B2921N, Sabbatical bench fee K800324N) and Ghent University (BOF-GOA 01G02124) for research funding. II and JL acknowledge Horizon Europe EIC Pathfinder program through project 101098649–UNICORN and IKUR Strategy under the collaboration agreement between Ikerbasque Foundation and BCMaterials on behalf of the Department of Education of the Basque Government.

## ABBREVIATIONS

BOE: Bloch Orbital Expansion; QD: Quantum Dot; DFT: Density Functional Theory; HOMO: Highest Occupied Molecular Orbital; LUMO: Lowest Unoccupied Molecular Orbital; MO: Molecular Orbital

**Table of Contents.**

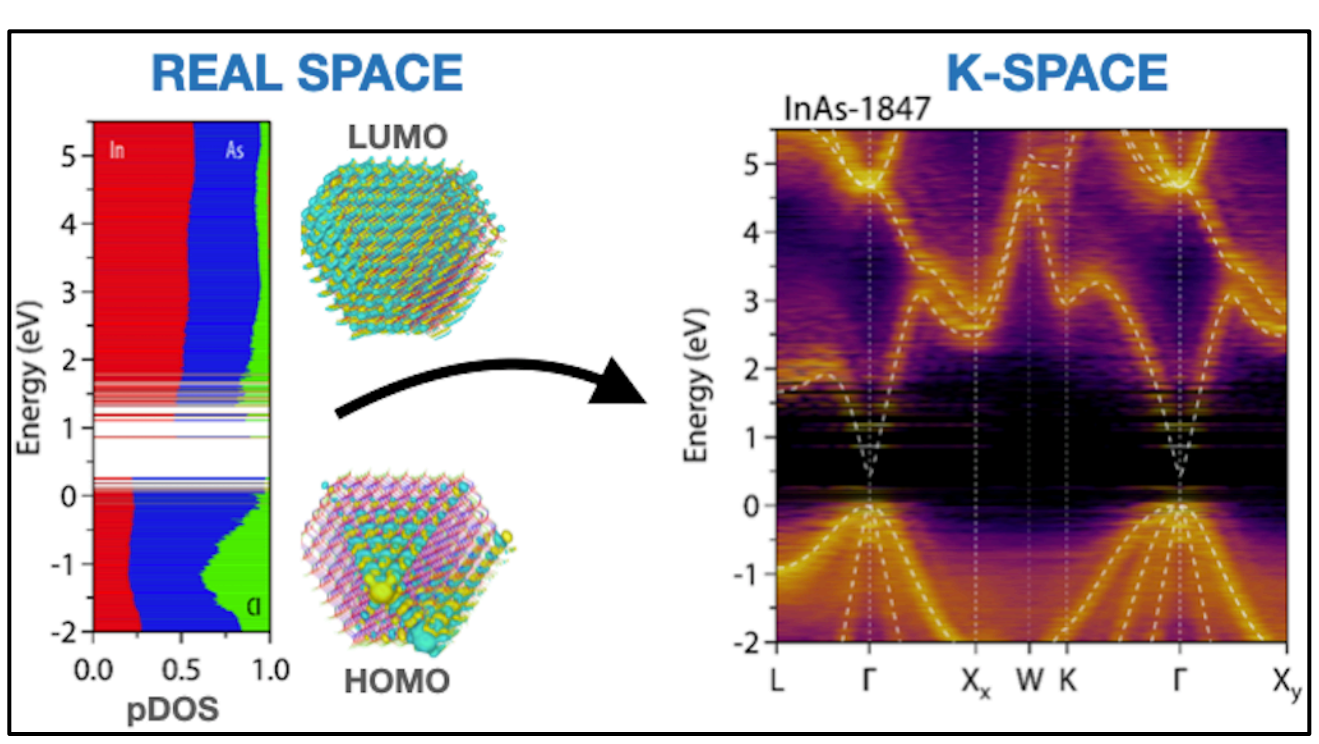
REAL SPACE
K-SPACE
LUMO
HOMO
InAs-1847
Energy (eV)
pDOS
In
As
Cl
L
Γ
$X_x$
W
K
Γ
$X_y$